\documentclass[Afour,sageh,times]{sagej}

\usepackage{url}
\usepackage{graphicx}
\usepackage{tikz}
\usetikzlibrary{patterns}
\usepackage{listings}
\lstdefinestyle{proto}{
    basicstyle=\scriptsize,
    frame=single,
    breaklines=true
}
\lstdefinestyle{example}{
    basicstyle=\scriptsize,
    language=C,
    frame=tb,
    showstringspaces=false,
    breaklines=true
}

\usepackage[hidelinks]{hyperref}

\newcommand\MPI[1]{\texttt{MPI\_\-#1}}
\newcommand\THREAD[1]{\texttt{MPI\_THREAD\_\-#1}}
\newcommand\MPIpThreads{MPI$+$Threads}
\newcommand\MPIxThreads{MPI$\times$Threads}

\begin{document}

\title{Designing and Prototyping Extensions to MPI in MPICH}

\date{}
\author{Hui Zhou\affilnum{1},  Ken Raffenetti\affilnum{1}, Yanfei Guo\affilnum{1},  Thomas Gillis\affilnum{1}, Robert Latham\affilnum{1},  Rajeev Thakur\affilnum{1}}

\affiliation{\affilnum{1}Argonne National Laboratory, Lemont, IL 60439, USA}

\corrauth{Hui Zhou, Argonne National Laboratory, Lemont, IL 60439, USA}
\email{zhouh@anl.gov}

\begin{abstract}
As HPC system architectures and the applications running on them continue to evolve, the MPI standard itself must evolve. 
The trend in current and future HPC systems toward powerful nodes with multiple CPU cores and multiple GPU accelerators makes efficient support for hybrid programming critical for applications to achieve high performance. 
However, the support for hybrid programming in the MPI standard has not kept up with recent trends. 
The MPICH implementation of MPI provides a platform for implementing and experimenting with new proposals and extensions to fill this gap and to gain valuable experience and feedback before the MPI Forum can consider them for standardization. 
In this work, we detail six extensions implemented in MPICH to increase MPI interoperability with other runtimes, with a specific focus on heterogeneous architectures.
First, the extension to MPI generalized requests lets applications integrate asynchronous tasks into MPI's progress engine.
Second, the iovec extension to datatypes lets applications use MPI datatypes as a general-purpose data layout API beyond just MPI communications.
Third, a new MPI object, MPIX\_Stream, can be used by applications to identify execution contexts beyond MPI processes, including threads and GPU streams. MPIX stream communicators can be created to make existing MPI functions thread-aware and GPU-aware, thus providing applications with explicit ways to achieve higher performance.
Fourth, MPIX Streams are extended to support the enqueue semantics for offloading MPI communications onto a GPU stream context. 
Fifth, thread communicators allow MPI communicators to be constructed with individual threads, thus providing a new level of interoperability between MPI and on-node runtimes such as OpenMP.
Lastly, we present an extension to invoke MPI progress, which lets users spawn progress threads with fine-grained control to adapt the communication performance to their application designs.
We describe the design and implementation of these extensions, provide usage examples, and highlight their expected benefits with performance results.
\end{abstract}

\keywords{Distributed Computing, Message Passing Interface, MPI+Threads, MPI+GPU, MPICH}

\maketitle
\section{Introduction}
The Message Passing Interface (MPI), first released in 1994, is an amazing feat in the history of computer science. It took merely several months from the conception to its initial draft and another several months for a voluntary forum consisting of collaborations from about 40 organizations to formally produce the MPI-1 standard.
Again beyond anyone's expectation,  MPI was quickly adopted and became the \textit{de facto} standard for distributed parallel computing.
Today the computer hardware that runs high-performance computing (HPC) applications is dramatically different from when MPI was first conceived, yet MPI remains one of the dominant runtimes. Most scientific applications still rely on MPI to reach state-of-the-art performance on leading computing facilities.

The longevity of MPI can be attributed to its well-balanced design that addresses both simplicity and composability on the user side and portability and performance on the implementation side.
The MPI model abstracts an explicit, collaborative, and synchronization-centric paradigm for parallel programs, which remains unique among alternative runtimes.
Application programmers can use MPI to engineer parallel algorithms that explicitly optimize for data locality and minimize synchronizations.
On the other hand, the abstraction leaves sufficient flexibility for implementations to adapt to the evolving hardware architectures.
For example, during the petascale push, computer architectures were shifting from a few cores per node to many cores per node, and the performance heavily depended on the optimal use of shared memory.
This situation challenged MPI's distributed process model.
MPI implementations had to be redesigned to take advantage of shared-memory-based communication; see \cite{nemesis2006}. However, MPI's core APIs were largely unchanged. Legacy MPI applications could reap the performance gain on newer systems without requiring significant code changes. Indeed,
\cite{krawezik2003} showed that it was more straightforward to obtain good performance in MPI than in alternative runtimes such as OpenMP.

Today, we are entering the period of exascale computing, and the computing hardware architecture is going through yet another dramatic shift. In particular, multiple graphics processing units (GPUs) per node have become common, and applications are increasingly relying on GPU offloading for performance.
While MPI is still the \textit{de facto} choice for internode communication, it is increasingly challenged by alternative runtimes for on-node programming.
Thus, hybrid programming models, such as MPI+Threads and MPI+GPU, are no longer avoidable.
The current MPI standard supports the hybrid programming model by allowing implementations to hide the complexity behind a configurable compatibility level. For example, with \THREAD{MULTIPLE}, multiple threads can call MPI functions concurrently as if they were called in a serial order. With GPU-aware MPI implementations, users can send MPI messages directly using GPU device memory.
While these conveniences allow developers to quickly port their applications into the hybrid model, however, it is tricky to maintain the communication performance relative to a pure MPI model.
The core issue is due to MPI not recognizing execution contexts other than an MPI process. When a user calls MPI from a thread context, for example, the MPI library is unable to reliably tell whether two calls will run into race conditions and thus requires thread-safety measures such as a global mutex. This communication serialization prevents good performance. The state-of-the-art implementations can avoid some of the serialization by gathering hints within the API semantics, but reliably achieving performance in \THREAD{MULTIPLE} remains tricky; see \cite{Rohit-21}.
Similarly, with GPU-aware MPI, it is tricky for implementations to optimize GPU resources and avoid extra memory registration overhead without the application's ability to directly tell MPI about its memory context.

Another challenge with hybrid programming models is the lack of interoperability between runtimes. For example, MPI+OpenMP applications use MPI for parallelization and communications between processes and use OpenMP for parallelization and communications between threads despite the similarity in the parallel semantics and algorithms. And between MPI and various asynchronous frameworks, separate progress models exist, which may interfere with each other and negatively impact performance.
Thus, hybrid programming models often introduce redundancy and inefficiency, and the situation can be addressed only if there is better interoperability between the runtimes. We need to make hybrid programming more homogeneous.

To keep up with the shifting HPC architecture, the MPI standard needs to be evolved.
However, evolving an established standard such as MPI is very challenging.
First, a large amount of legacy code cannot afford a significant rewrite.
Thus, the development of a new MPI standard has to be backward-compatible.
Second, the new APIs need to follow a consistent design so that they can be readily adopted by existing MPI projects.
Third, the new APIs need to address the key challenges.
Fourth, the new APIs need to be sufficiently abstract so that they will remain applicable to future architectures and technology.
These criteria are important to maintain the success and longevity of MPI.
We cannot achieve perfect new APIs simply by design. Successful APIs require experimentation and collection of feedback.
Many message-passing frameworks existed before MPI, including PARMACS (\cite{PARMACS}), Zipcode (\cite{Zipcode}), Chimp (\cite{CHIMP}), PVM (\cite{PVM}), Chameleon (\cite{chameleon}), and PICL (\cite{PICL}). The feedback from these early frameworks provided the foundation behind the initial success of the MPI standard.

In MPICH, we believe experimentation is crucial for continuously evolving MPI. As one of the main MPI implementations, MPICH can reach a wide user base and is thus best positioned to collect user experiences from experimental functionality, in addition to implementation experience from the developer team itself.
Experimental APIs in MPICH use the \texttt{MPIX} prefix to differentiate from the official standard APIs, which use the \texttt{MPI} prefix.

In this paper, we provide an overview of major extensions that are available in the latest MPICH-4.2.0 release. The extensions are grouped by features. They are (1) generalized requests, (2) derived datatypes, (3) mpix streams, (4) offloading asynchronous operations, (5) thread communicators, and (6) general progress. In each section we discuss the background on the rationale of the extensions, listing the API function prototypes and documenting their usage; we also provide example code to illustrate how to use these extensions, and we provide testing results and discuss their benefits.

Most of these extensions are outcomes from the research effort funded by the Exascale Computing Project. They address the key performance challenges and the interoperability issues faced by modern HPC applications to adapt to modern computing environments.
We plan to propose the extensions for consideration for the next version of the MPI specification. Meanwhile, they are fully available in the current MPICH release, and they are expected to be available in most MPICH-derived vendor MPIs such as Cray MPICH.
In this paper, we try to provide more comprehensive documentation for these experimental APIs in the hope of seeing more adoption from the HPC community. The experience and feedback will help us and the MPI Forum propose more sound proposals for the new MPI standard.

\section{Generalized Requests}
\subsection{Background}
One of the notable aspects of MPI is that its design considered external interfaces and extensibility early on.
For example, the profiling interface has been specified since MPI-1 to ensure that authors of profiling tools can interface their codes relatively easily to any MPI function.
Another example is generalized requests, added in MPI-2, to allow custom asynchronous tasks to be layered on top of MPI. One can start nonblocking communications and multiple non-MPI asynchronous operations---for example, I/O offloading tasks---and then synchronize all with a single \texttt{MPI\_Waitall}.

\begin{figure}[htp]
    \centering
    \includegraphics[clip,width=\columnwidth]{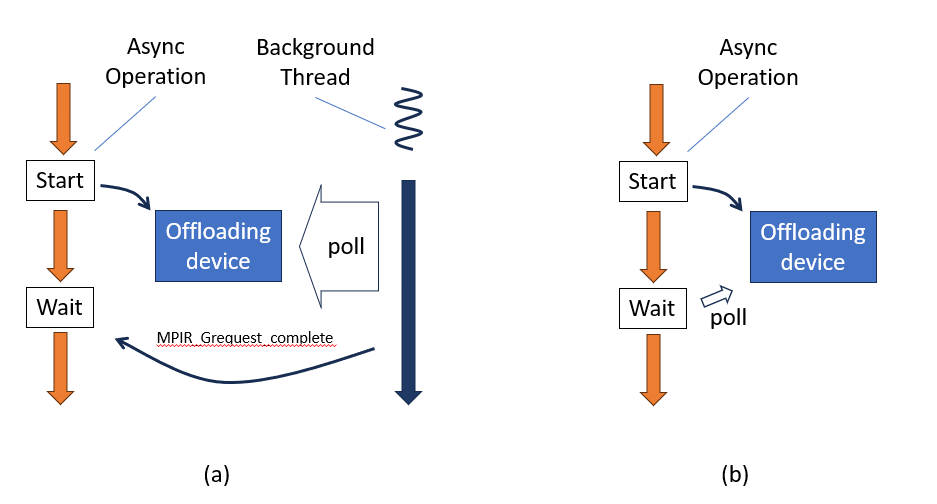}
    \caption{Diagrams illustrating asynchronous operations via MPI generalized request. (a) The current standard API requires background threads to complete the request. (b) Extension may eliminate the need for a background thread.}
    \label{fig:grequest}
\end{figure}

However, the current standard omits the progress mechanism for generalized requests. It is assumed that programs that use generalized requests will launch separate threads to do the progress and complete the generalized requests outside MPI, as illustrated in Figure \ref{fig:grequest}a. In practice, most asynchronous frameworks have their own progress models and often do not require active progress. With the current specification of generalized requests, an extra progress thread is always needed because the external asynchronous runtimes do not interoperate with MPI and have no mechanisms to complete a generalized MPI request. For example, in the case of asynchronous I/O, the operating system manages the completion of I/O operations. In the case of GPU offloading, the GPU runtimes manage the launching of kernels and synchronization of completion events. Since the operating system and GPU runtimes do not know how to complete a generalized request, users still need to launch separate polling mechanisms to just complete the generalized requests after the underlying tasks are completed.
This makes the benefit of using a single \texttt{MPI\_Waitall} to complete all tasks unattractive.

To address this shortcoming, MPICH provides an extension to allow user programs to attach a poll function callback to a generalized request, eliminating the need for separate completion mechanisms. This simplifies the usage of generalized requests, as illustrated in Figure \ref{fig:grequest}b.

In addition to the poll function callback, applications may benefit from waiting for asynchronous tasks to complete rather than repeatedly calling the poll functions for each request. This is achieved by supplying a wait function callback that accepts an array of asynchronous tasks.

This extension is used by ROMIO, an MPI-IO implementation used by both MPICH and OpenMPI. For reference, see \cite{latham2007extending}.

\subsection{Extension}
\begin{lstlisting}[style=proto]
int MPIX_Grequest_start(
    MPI_Grequest_query_function *query_fn,
    MPI_Grequest_free_function *free_fn,
    MPI_Grequest_cancel_function *cancel_fn,
    MPIX_Grequest_poll_function *poll_fn,
    MPIX_Grequest_poll_function *wait_fn,
    void *extra_state,
    MPI_Request *request)                 
\end{lstlisting}
The call starts a generalized request and returns a handle to it in \texttt{request}. 
The syntax and meaning of the callback functions \texttt{query\_fn}, \texttt{free\_fn}, and \texttt{cancel\_fn} are the same as in the standard \texttt{MPI\_Grequest\_start}. 
The callback \texttt{poll\_fn} and \texttt{wait\_fn} are specified below.
\begin{lstlisting}[style=proto]
typedef int MPIX_Grequest_poll_fn(void *extra_state,
                                  MPI_Status *status); 

typedef int MPIX_Grequest_wait_fn(int count,
                                  void *array_of_states,
                                  double timeout,
                                  MPI_Status *status);                                   
\end{lstlisting}

If the underlying asynchronous task already has a completion mechanism, the poll function just needs to query the status and complete the generalized request if the underlying task is completed.

It is unspecified when the \texttt{poll\_fn} will be called or how often it will be called. The intention is to leave flexibility for implementations to optimize for performance.
Assuming \texttt{poll\_fn} will eventually complete the generalized request, calls to \texttt{MPI\_Wait} will eventually return, and repeated calls to \texttt{MPI\_Test} will eventually complete the request.

\subsection{Example}
The following example in CUDA illustrates the usage of the generalized request extension with \texttt{poll\_fn} callback to wrap an asynchronous CUDA task.
For simplification, \texttt{query\_fn}, \texttt{free\_fn}, and \texttt{cancel\_fn} are left as empty functions. \texttt{wait\_fn} is set to \texttt{NULL} to omit the wait optimization. In \texttt{poll\_fn}, we query the CUDA event and call \texttt{MPI\_Grequest\_complete} once the task is completed.
\begin{lstlisting}[style=example]
/* grequest.cu */
#include <mpi.h>
#include <stdio.h>

const size_t N = 1000000;

struct grequest_state {
    cudaEvent_t event;
    MPI_Request request;
};

__global__
void saxpy(int n, float a, float *x, float *y)
{
  int i = blockIdx.x*blockDim.x + threadIdx.x;
  if (i < n) y[i] = a * x[i] + y[i];
}

int query_fn(void *extra_state, MPI_Status *status)
    {return MPI_SUCCESS;}
int free_fn(void *extra_state)
    {return MPI_SUCCESS;}
int cancel_fn(void *extra_state, int complete)
    {return MPI_SUCCESS;}

int poll_fn(void *extra_state, MPI_Status *status)
{
    struct grequest_state *p = (struct grequest_state *) extra_state;
    if (cudaEventQuery(p->event) == cudaSuccess) {
        MPI_Grequest_complete(p->request);
    }
    return MPI_SUCCESS;
}

int main(int argc, char **argv)
{
    MPI_Init(NULL, NULL);

    struct grequest_state state;
    cudaEventCreate(&state.event);
    MPIX_Grequest_start(query_fn, free_fn, cancel_fn, poll_fn, NULL, &state, &state.request);

    float a, *x, *y, *d_x, *d_y;
    x = (float*)malloc(N*sizeof(float));
    y = (float*)malloc(N*sizeof(float));
    cudaMalloc(&d_x, N*sizeof(float));
    cudaMalloc(&d_y, N*sizeof(float));

    a = 2.0f;
    for (int i = 0; i < N; i++) {
        x[i] = 1.0f;
        y[i] = 2.0f;
    }
    cudaMemcpyAsync(d_x, x, N*sizeof(float), cudaMemcpyHostToDevice, NULL);
    cudaMemcpyAsync(d_y, y, N*sizeof(float), cudaMemcpyHostToDevice, NULL);
    saxpy<<<(N+255)/256, 256, 0, NULL>>>(N, a, d_x, d_y);
    cudaMemcpyAsync(y, d_y, N*sizeof(float), cudaMemcpyDeviceToHost, NULL);
    cudaEventRecord(state.event, NULL);

    MPI_Wait(&state.request, MPI_STATUS_IGNORE);

    cudaFree(d_x);
    cudaFree(d_y);
    cudaEventDestroy(&state.event);
    free(x);
    free(y);

    MPI_Finalize();

    return 0;
}
\end{lstlisting}

\section{Derived Datatypes}
\begin{figure}
    \includegraphics[width=\columnwidth]{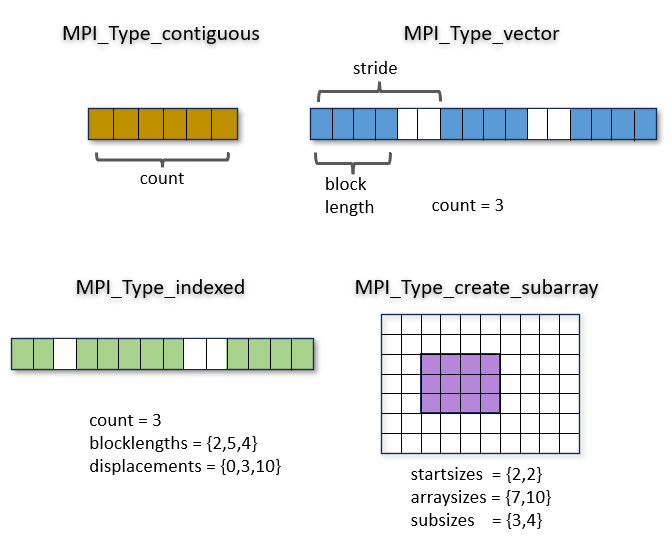} 
    \caption{Some MPI derived datatypes with illustrations of their creation routines.}
    \label{fig:datatypes}
\end{figure}

MPI datatype is an efficient method to describe a data layout. Figure~\ref{fig:datatypes} illustrates several examples. If we assume a 3-dimensional array of $N_x\times N_y \times N_z$, its most fragmented surface---the $YZ$ surface---will have $N_y \times N_z$ non-contiguous segments. Listing all the segments as an iovec array will require  $O(N_yN_z)$ both in memory and in time. Using MPI datatype, the layout is a two-level nested strided vector. Thus it takes constant cost regardless of the actual number of segments.
MPI datatypes can be arbitrarily nested, and they support offsets. Even a non-contiguous layout with overlapping segments can be described.
The abstraction of MPI datatypes allows clean interface designs that focus on logic in data operations rather than the subtleties in data layout. 

The expressive power of MPI datatypes and their efficiency against brute-force segment listing are generally useful. In the current standard, however, MPI datatypes are opaque objects that are useful only in MPI communications. There is no easy way to retrieve the segments in an MPI datatype from outside an MPI library, thus preventing MPI datatypes from being used in wider applications.

Non-contiguous data access is a common pattern in HPC applications. To facilitate general usage of MPI datatypes,
MPICH provides an extension to let users randomly query the segment (as iovec) of a datatype, thus allowing utilities and libraries to be built directly using MPI datatypes rather than reinventing another layout description scheme.
The extension makes MPI datatypes interoperable with codes that employ iovec-based algorithms.

\subsection{Extension}
\begin{lstlisting}[style=proto]
int MPIX_Type_iov_len(MPI_Datatype datatype,
                      MPI_Count max_iov_bytes,
                      MPI_Count *iov_len,
                      MPI_Count *actual_iov_bytes)
\end{lstlisting}
This function queries an MPI datatype for the number of \texttt{iov} segments in \texttt{iov\_len} within \texttt{max\_iov\_bytes}.
If \texttt{max\_iov\_bytes} is greater than the total size of the datatype---returned from \texttt{MPI\_Type\_size}---or $-1$, \texttt{iov\_len} is the total number of \texttt{iov} segments in the datatype.
Otherwise, it returns the number of whole segments inside \texttt{max\_iov\_bytes}.
\texttt{actual\_iov\_bytes} returns the number of bytes of the first \texttt{iov\_len} segments. 

Typically this function is called with \texttt{max\_iov\_bytes} equal to $-1$ to query the total number of segments. But \texttt{max\_iov\_bytes} can be used to bisect the byte offset of an arbitrary segment.
\begin{lstlisting}[style=proto]
int MPIX_Type_iov(MPI_Datatype datatype,
                  MPI_Count iov_offset,
                  MPIX_Iov *iov, 
                  MPI_Count max_iov_len,
                  MPI_Count *actual_iov_len)                      
\end{lstlisting}
This function returns a list of \texttt{iov} segments from \texttt{iov\_offset} up to \texttt{iov\_offset} $+$ \texttt{max\_iov\_len}.
\texttt{actual\_iov\_len} returns the actual number of segments returned.

\texttt{MPIX\_Iov} is defined to be compatible with \texttt{struct iovec} in the standard C library.
\begin{lstlisting}[style=proto]
typedef struct MPIX_Iov {
    void *iov_base;
    MPI_Aint iov_len;
} MPIX_Iov;
\end{lstlisting}

\subsection{Example}
The following example illustrates the usage of the datatype iovec extensions.
We create a datatype using \texttt{MPI\_Type\_create\_subarray} that represents the layout of a subportion of a 3D volume. Then we use the extension routines to query the segment information.
\begin{lstlisting}[style=example]
/* typeiov.c */
#include <mpi.h>
#include <stdio.h>
#include <limits.h>

struct value {
    double a;
    double b;
};

int main(void)
{
    MPI_Init(NULL, NULL);

    MPI_Datatype value_type, volume_type;
    /* Create value type as a contiguous blob */
    MPI_Type_contiguous(sizeof(struct value), MPI_BYTE, &value_type);

    /* Create a sub-volume inside a 3-dimensional array */
    int full_sizes[3] = {1000, 1000, 1000};
    int sub_sizes[3] = {100, 100, 100};
    int sub_offsets[3] = {300, 300, 300};
    MPI_Type_create_subarray(3, full_sizes, sub_sizes, sub_offsets,
                             MPI_ORDER_C, value_type, &volume_type);
    MPI_Type_commit(&volume_type);

    /* Assess the segment informantion using the IOV extensions */
    MPI_Count iov_len, iov_bytes;
    MPIX_Type_iov_len(volume_type, INT_MAX, &iov_len, &iov_bytes);
    printf("iov_len = %ld, iov_bytes = %ld\n", (long) iov_len, (long) iov_bytes);

    MPIX_Iov iov[4];
    MPI_Count actual_iov_len;
    MPIX_Type_iov(volume_type, 0, iov, 4, &actual_iov_len);
    for (int i = 0; i < 4; i++) {
        printf("iov[%d]: %p - %ld\n", i, iov[i].iov_base, (long) iov[i].iov_len);
    }

    MPI_Type_free(&volume_type);
    MPI_Type_free(&value_type);
    MPI_Finalize();
    return 0;
}
\end{lstlisting}

\section{MPIX Streams}
\subsection{Background}
The current trend in HPC computing is to deploy applications using a hybrid MPI+X model,
where MPI is the programming model for the internode communication, and
X refers to an on-node programming model such as OpenMP or an accelerator runtime such as CUDA.
This corresponds to the trend in hardware architecture where a typical node consists of many CPU cores
and several offloading accelerators with mixed types of shared memory.
Making MPI and the X runtime cooperate is a challenge but it is also crucial to unlock the maximum performance potential of the system.

The current support for MPI+X is mainly at compatibility. Starting with the threading context,
MPI supports four thread-compatibility levels: \THREAD{SINGLE},
\THREAD{FUNNELED}, \THREAD{SERIALIZED}, and \THREAD{MULTIPLE}.
When the appropriate thread level is chosen, threaded applications can work
correctly with MPI without MPI specifically acknowledging the runtime.
Similarly, the current support for MPI+GPU is for MPI implementations to be GPU aware. Recent
MPICH, MVAPICH, and Open MPI releases are all able to detect GPU buffers without hints from users and
make MPI work without a GPU-specific MPI interface.
Currently, the GPU compatibility level is assumed.
With MPICH, the environment variable \texttt{MPIR\_CVAR\_ENABLE\_GPU} can be used to switch on or off GPU compatibility. The default is on if the library is built with GPU support.

With MPI+Threads, while it is successful on the compatibility side, the
performance side has been a multi-decade struggle.
It is notorious that an application with latency-sensitive communication patterns using \THREAD{MULTIPLE} is likely to meet with dismal performance.
The cause of this performance penalty is well known. It comes from the critical sections introduced by MPI
communications and contentions between multiple threads.
Much research has been done on both the application side (see \cite{wang2019multi})
and the implementation side (see \cite{amer2015,OMPICRI-19,Rohit-20}) to address
the performance of MPI+Threads. To reach good performance, applications need to
make sure that the communications can happen in parallel, in other words, logically concurrent; and the implementations
need to map the communications to separate communication channels to allow the
communication to proceed in parallel.
Without an explicit MPI interface,  making the latter mapping to match the application layer concurrency remains an art. 
A mismatch will result in either incorrect results or extra thread contention and bad performance.
The underlying communication channels are conventionally referred to as network endpoints.
In MPICH, network endpoints are abstracted into virtual communication interfaces (VCIs). See \cite{Rohit-20, Rohit-21}. 

The performance story of MPI+GPUs is different from that of MPI+Threads but shares a similar cause -- MPI lacks an explicit interface for the GPU execution context.
Accelerators typically require special runtime to coordinate between CPU and accelerator
executions. The launching and synchronization between CPU context and
accelerator context are carried out by the accelerator runtime. A key
performance factor here is how to minimize the launching and synchronization
cost. To optimize the performance, we need MPI operations to be enqueued 
to an accelerator execution context and then let the accelerator runtime 
manage its actual execution. In order to realize this new mode of MPI
operations, new MPI interfaces that work directly with accelerator execution
context are needed.

\begin{figure}
    \includegraphics[width=\columnwidth]{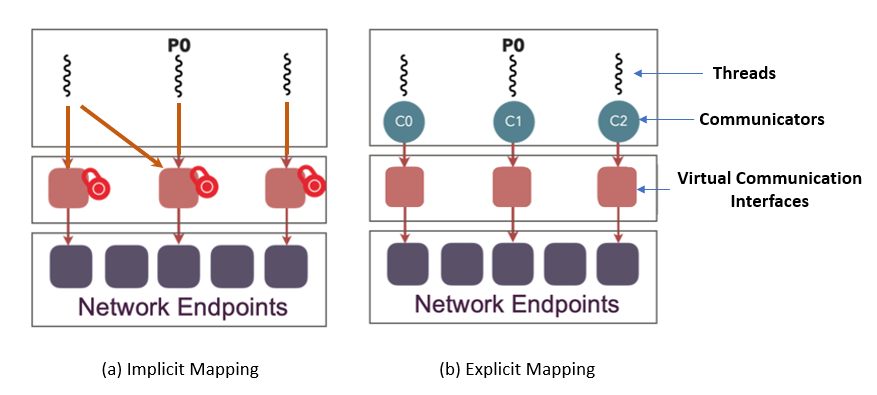}
    \caption{Diagram illustrating mapping communications to network endpoints. (a) Implicit scheme maps communications implicitly to internal virtual communication interfaces, requires locking, and may result in mismapping. (b) Explicit scheme requires explicit context from communicators and may eliminate locking.}
    \label{fig:mpixstreams}
\end{figure}

Although the implicit strategy to obtain good performance has been demonstrated successfully in the past, it is far from consistent.
It often relies on application programmers to engineer code patterns and provide unofficial hints to allow MPI implementations to correctly match the contexts and intentions.
Often, it is simpler for the application to explicitly pass the contexts into the MPI library so it can perform according to users' expectation.
This explicit strategy will require new extensions since the current standard does not have a concept to represent execution contexts other than MPI processes.
This new extension is named MPIX Stream. For reference, see \cite{mpix-stream}.
The implicit mapping and explicit mapping of communications to network endpoints are illustrated in Figure~\ref{fig:mpixstreams}.

\subsection{Extension}
MPICH introduces an abstract object, MPIX stream, to facilitate a direct mapping from
user-level runtime execution contexts to MPI's internal execution contexts or VCIs.
\begin{lstlisting}[style=proto]
int MPIX_Stream_create(MPI_Info info,
                       MPIX_Stream *stream)
  
int MPIX_Stream_free(MPIX_Stream *stream)
\end{lstlisting}
To MPI, an MPIX stream represents a local serial execution context.
Any runtime execution contexts outside MPI, as long as the serial semantics is strictly followed, can be
associated with an MPIX stream. Examples include kernel threads, user-level threads,
GPU queuing streams, or even code across multiple threads with serialized synchronizations.

Info hints can be used to create implementation-supported special streams, for
example, a CUDA stream. Otherwise, \texttt{MPI\_INFO\_NULL} can be used.
Whether the returned stream is backed by hardware network endpoints is implementation-dependent.
By default, MPICH will try to allocate distinct network endpoints for each new stream and return failure if it runs out of available endpoints.
This affords the application with a predictable performance. 
With unique endpoints and the strict serial execution context guaranteed by the stream context, the implementation can safely skip critical sections and thus achieve maximum parallel performance.
For streams representing GPU streams, MPICH may reuse network endpoints. The reason is that the asynchronous nature of GPU streams makes explicit isolation to communication traffic less critical. GPU-representing streams are used for offloading context and will be covered in the next section.

Because network endpoints are finite resources, users should free the
streams to make the resource available for future allocation.

To apply MPIX stream, one creates stream communicators with local stream context attached.
\begin{lstlisting}[style=proto]
int MPIX_Stream_comm_create(MPI_Comm comm,
                            MPIX_Stream stream,
                            MPI_Comm *newcomm) 
\end{lstlisting}
Just like other communicator creation functions, this call is collective over all processes in the old communicator \texttt{comm}.
Note that the stream is local to each process. The stream communicator is a collection of local execution contexts.
Once the stream communicator is created, users can make communication calls the same way as using a normal communicator.
No additional adaptation from the user code is needed. The MPI library, on the other hand, can check and use the correctly mapped network endpoints and skip any thread-safety measures, thus achieving the best potential parallel performance.

Not all processes have to attach a local MPIX stream. \texttt{MPIX\_STREAM\_NULL} can be used to denote no stream, or an implicit scheme can be used for that process.
It is valid to create a stream communicator with all processes specifying \texttt{MPIX\_STREAM\_NULL}, in which case the new stream communicator will revert to the same behavior as a conventional communicator created from \texttt{MPI\_Comm\_dup}.

When the stream communicator is no longer needed, it should be freed by using \texttt{MPI\_Comm\_free}.

The stream communicator created by using \texttt{MPI\_Stream\_comm\_create} is sometimes referred to as a single-stream communicator since each process has only a single stream attached. This is convenient but restrictive, since a local stream can  communicate only with a fixed stream in the remote process. Sometimes a process may need to communicate with multiple remote streams. For example, an event dispatch system may have a listening process serving arbitrary events issued from any remote contexts. Since a single-stream communicator fixes the remote context, multiple single-stream communicators are needed.
This is inconvenient. In addition, wildcard receives cannot be issued across multiple communicators.
To overcome this restriction, user can create a multiple-stream communicator, in which each process can be attached with an array of local streams.
\begin{lstlisting}[style=proto]
int MPIX_Stream_comm_create_multiplex(
    MPI_Comm comm,
    int count,
    MPIX_Stream array_of_streams[],
    MPI_Comm *newcomm)
\end{lstlisting}
Because each process has multiple streams attached, when issuing communication calls, extra parameters are needed to specify which local and remote stream to be used.
\begin{lstlisting}[style=proto]
int MPIX_Stream_send(const void *buf, int count,
                     MPI_Datatype datatype,
                     int dest, int tag, 
                     MPI_Comm comm,
                     int source_stream_index,
                     int dest_stream_index)

int MPIX_Stream_isend(const void *buf, int count,
                      MPI_Datatype datatype,
                      int dest, int tag,
                      MPI_Comm comm,
                      int source_stream_index,
                      int dest_stream_index,
                      MPI_Request *request)

int MPIX_Stream_recv(void *buf, int count,
                     MPI_Datatype datatype,
                     int source, int tag,
                     MPI_Comm comm,
                     int source_stream_index,
                     int dest_stream_index,
                     MPI_Status *status)

int MPIX_Stream_irecv(void *buf, int count,
                      MPI_Datatype datatype,
                      int source, int tag,
                      MPI_Comm comm,
                      int source_stream_index,
                      int dest_stream_index,
                      MPI_Request *request)
\end{lstlisting}
In \texttt{MPI\_Stream\_recv} and \texttt{MPI\_Stream\_irecv}, $-1$ can be used in \texttt{source\_stream\_index} to specify an any-stream receive.

The attached stream can be retrieved by using \texttt{MPI\_Comm\_get\_stream}.
\begin{lstlisting}[style=proto]
  int MPIX_Comm_get_stream(MPI_Comm comm, int idx,
                           MPIX_Stream *stream)
\end{lstlisting}

\subsection{Example}
The following example shows how MPIX streams are used in a \texttt{MPI\_THREAD\_MULTIPLE} program. Multiple threads from two processes form communication pairs.
Using MPIX streams, each pair communicates on separate streams, thus are semantically concurrent.
\begin{lstlisting}[style=example]
#define NT 4

int main(void) {
    int rank;
    int tl;
    MPI_Init_thread(NULL, NULL, MPI_THREAD_MULTIPLE, &tl);
    MPI_Comm_rank(MPI_COMM_WORLD, &rank);

    MPIX_Stream streams[NT];
    MPI_Comm comms[NT];
    for (int i = 0; i < NT; i++) {
        MPIX_Stream_create(MPI_INFO_NULL, &streams[i]);
        MPIX_Stream_comm_create(MPI_COMM_WORLD, streams[i], &comms[i]);
    }

    #pragma omp parallel num_threads(NT)
    {
        int id = omp_get_thread_num();
        char buf[100];
        int tag = 0;
        if (rank == 0) {
            MPI_Send(buf, 100, MPI_CHAR, 1, tag, comms[id]);
        } else if (rank == 1) {
            MPI_Recv(buf, 100, MPI_CHAR, 0, tag, comms[id], MPI_STATUS_IGNORE);
        }
    }
    for (int i = 0; i < NT; i++) {
        MPIX_comm_free(&comms[i]);
        MPIX_Stream_free(&streams[i]);
    }

    MPI_Finalize();
    return 0;
}   
\end{lstlisting}

\subsection{Evaluation}
\begin{figure}
    \includegraphics[width=\columnwidth]{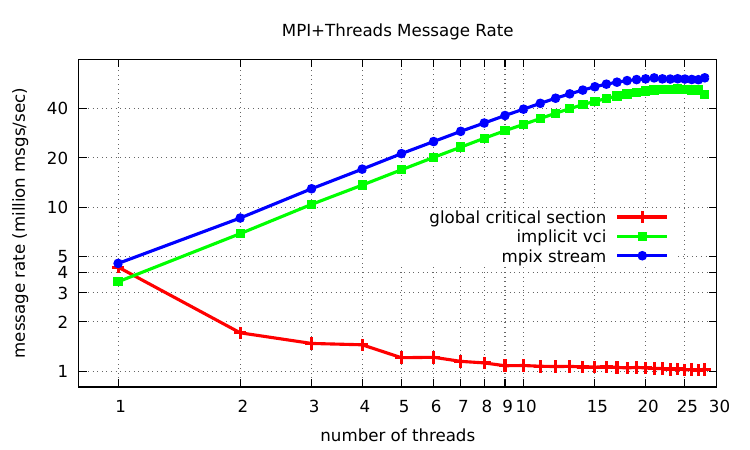}
    \caption{Multithread message rate on 8-byte messages using
    MPI\_Isend/MPI\_Irecv. The message rate using MPIX\_Stream is around $20\%$ higher than with implicit VCIs.}
    \label{fig:msgrate}
\end{figure}

To showcase the performance advantages of integrating MPIX stream into an MPI+Threads application, we carried out a microbenchmark test 
on an Intel Skylake cluster in the Joint Laboratory for System Evaluation at Argonne
National Laboratory. 
In this test, we have used two nodes connected by Mellanox InfiniBand (EDR-IB).
The microbenchmark launches a number of threads for each process, each sending 8-byte messages to a corresponding thread in another process.
In Figure \ref{fig:msgrate}, we measure the message rate in three different configurations.

First, the red curve shows the message rate with MPICH configured to use the global
critical section. This is the default configuration in earlier MPICH releases (before version 4.0).
We observe that the total message rate drops as soon as more threads start to compete for the critical section.
A similar performance decrease might occur even in the absence of a global critical section if multiple threads are assigned to a single internal network channel, leading to contention among them.

Second, the green curve shows the message rate when MPICH is configured to utilize the per-VCI critical section (default in the current release of MPICH). In this configuration, MPICH implicitly hashes communications to various VCIs. The microbenchmark is tailored to achieve perfect implicit hashing, resulting in good scaling as we increase the number of threads.
It is worth noting that the message rate with a single thread is lower than the corresponding rate observed with the global critical section. This discrepancy arises from the finer granularity of the per-VCI critical sections, necessitating multiple critical sections along the communication path, notably impacting both the receive path and progress engine. Even in the absence of contention, the additional locking and unlocking processes adversely affect performance.

Finally, the blue curve represents the benchmark rewritten to incorporate the stream communicators outlined in this section. In this setup, each stream communicator is linked to a distinct MPIX stream object per thread. Leveraging the semantics of MPIX stream, which ensures a serial execution context, our implementation eliminates the need for locking entirely. This optimization yields an approximate 20\% increase over the implicit scheme in the total message rate up to 20 concurrent threads.

\section{Offloading Asynchronous Operations}
\subsection{Background}
GPU execution contexts such as CUDA are offloading contexts. Operations are  issued only within the CPU context but are executed asynchronously by an offloading device, \textit{e.g.}~a GPU.
The offloaded tasks can be executed concurrently. To enable concurrency, programmers need specify which tasks cannot be executed concurrently, in other words, specify the dependencies between tasks.
Specifying the full dependency graph adds complexity to both the programming and the execution in the GPU runtimes. An alternative is to use a stream. In CUDA, this is a CUDA stream (\cite{gomez2012}).
A CUDA stream assumes sequential dependency for the tasks issued on the same stream. Tasks issued on different streams can be safely executed concurrently.

\begin{figure}
    \includegraphics[width=\columnwidth]{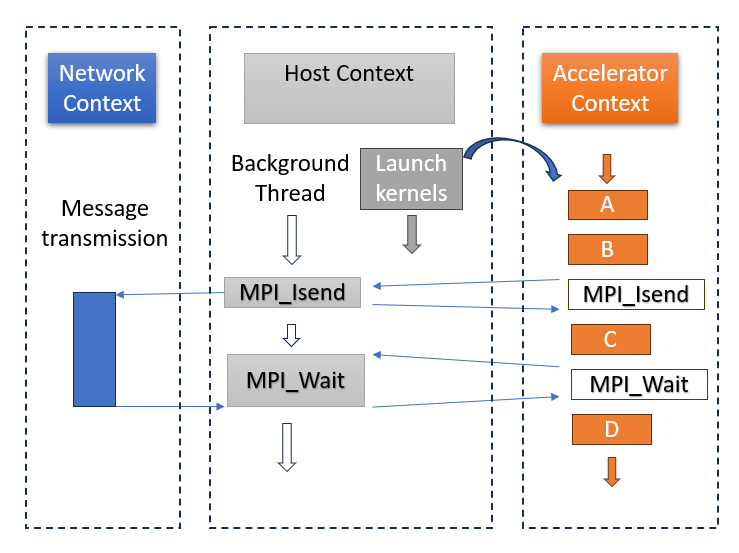}
    \caption{Diagram illustrating how MPI operations are launched into the accelerator context and triggered to run under the host context.}
    \label{fig:enqueue}
\end{figure}

An MPIX stream can be extended to represent an offloading stream such as a CUDA stream. To create such an offloading stream, one uses \texttt{MPIX\_Create\_stream} and specifies the offloading stream parameter via the \texttt{info} parameter.
The usage is similar to that of non-offloading streams. First, stream communicators are created by using \texttt{MPIX\_Stream\_comm\_create} and then  MPI communications are issued by using the same standard syntax, for example, \texttt{MPI\_Send}.
Because the attached stream is an offloading stream, the issued operation does not get executed or even started within the CPU execution context. Rather, operations will be queued and executed within the GPU stream context. The embedding of MPI communications into a GPU stream is illustrated in Figure~\ref{fig:enqueue}. 

While the offloading semantics is new to MPI, by creating stream communicators from GPU stream backed MPIX streams, we have the benefit of imposing the offloading semantics to existing MPI operations. Alternatively, \cite{Naveen-22} proposed a set of new enqueue APIs to achieve a similar offloading semantics. Their paper described only the send and receive operations. However, extensions based on MPIX streams can be readily extended to collectives and remote memory operations.

\subsection{Extension}
To create an MPIX stream representing an offloading GPU stream, we need pass the GPU stream object in \texttt{MPIX\_Create\_stream} using info hints.
However, an \texttt{MPI\_Info} object  supports values only as strings. A GPU queuing object
not only is not a string but is often opaque to the user. For example, is a CUDA
stream type, \texttt{cudaStream\_t}, an integer or a pointer, or could it be neither? To pass an opaque
binary as a string, MPICH provides an extension to set info hints with binary value.
\begin{lstlisting}[style=proto]
int MPIX_Info_set_hex(MPI_Info info, const char *key,
                      void *value, int vallen)
\end{lstlisting}
The following example code creates an MPIX stream representing a CUDA stream and then creates a stream communicator attaching the stream context.
\begin{lstlisting}[style=example]
  MPI_Info info;
  MPI_Info_create(&info);
  MPI_Info_set(info, "type", "cudaStream_t");
  MPIX_Info_set_hex(info, "value", &stream, sizeof(stream));

  MPIX_Stream mpi_stream;
  MPIX_Stream_create(info, &mpi_stream);

  MPI_Comm stream_comm;
  MPIX_Stream_comm_create(MPI_COMM_WORLD, mpi_stream, &stream_comm);
\end{lstlisting}

To issue MPI send and receive operations on this stream communicator, one can use \texttt{MPI\_Send} and \texttt{MPI\_Recv}.
Because of the offloading context, however, the communication gets enqueued to the GPU stream rather than getting executed immediately.
This can be surprising since there is no difference in code from regular MPI send and receive.
To make the enqueuing semantics explicit, MPICH provides the following aliases.
\begin{lstlisting}[style=proto]
int MPIX_Send_enqueue(const void *buf, int count,
                      MPI_Datatype datatype,
                      int dest, int tag,
                      MPI_Comm comm)

int MPIX_Recv_enqueue(void *buf, int count,
                      MPI_Datatype datatype,
                      int source, int tag,
                      MPI_Comm comm,
                      MPI_Status *status)
\end{lstlisting}
Both are merely aliases to \texttt{MPI\_Send} and \texttt{MPI\_Recv}, but they make the enqueuing semantics explicit. We highly recommend using these alias functions to issue enqueuing operations.

The non-blocking versions of send and receive are also provided.
\begin{lstlisting}[style=proto]
int MPIX_Isend_enqueue(const void *buf, int count,
                       MPI_Datatype datatype,
                       int dest, int tag,
                       MPI_Comm comm,
                       MPI_Request *request)

int MPIX_Irecv_enqueue(void *buf, int count,
                       MPI_Datatype datatype,
                       int source, int tag,
                       MPI_Comm comm,
                       MPI_Request *request)

int MPIX_Wait_enqueue(MPI_Request *request,
                      MPI_Status *status);

int MPIX_Waitall_enqueue(int count,
                         MPI_Request array_of_requests[],
                         MPI_Status *array_of_statuses)
\end{lstlisting}

One may be confused as to why we need non-blocking communication for the offloading context since the offloading context is already asynchronous.
It is helpful to recognize that there are three different contexts. One is the offloading context, which is asynchronous from the CPU context. The second is the starting and waiting for completion of the communications. This happens on the CPU and can be asynchronous from the communication itself. The actual communication is carried out by the networking device. The MPI non-blocking operation applies to the latter two contexts independent of the offloading context.

\subsection{Example}
The following example is an MPI+CUDA program illustrating the usage of the enqueuing extensions. Process 0 generates a portion of the data and sends it to process 1, which launches the kernel to do the computation after receiving the data. All memory copies, MPI send/receive, and computation kernels are asynchronously launched to a user-supplied CUDA stream. Note that \texttt{cudaStreamSynchronize} is completely avoided. 
\begin{lstlisting}[style=example]
/* enqueue.cu */
const float a_val = 2.0;
const float x_val = 1.0;
const float y_val = 2.0;

__global__
void saxpy(int n, float a, float *x, float *y)
{
  int i = blockIdx.x*blockDim.x + threadIdx.x;
  if (i < n) y[i] = a_val * x[i] + y[i];
}

int main(void)
{
    cudaStream_t stream;
    cudaStreamCreate(&stream);

    int rank;
    MPI_Init(NULL, NULL);
    MPI_Comm_rank(MPI_COMM_WORLD, &rank);

    float *x, *y, *d_x, *d_y;

    MPI_Info info;
    MPI_Info_create(&info);
    MPI_Info_set(info, "type", "cudaStream_t");
    MPIX_Info_set_hex(info, "value", &stream, sizeof(stream));

    MPIX_Stream mpi_stream;
    MPIX_Stream_create(info, &mpi_stream);

    MPI_Info_free(&info);

    MPI_Comm stream_comm;
    MPIX_Stream_comm_create(MPI_COMM_WORLD, mpi_stream, &stream_comm);

    /* Rank 0 sends x data to Rank 1, Rank 1 performs a * x + y and checks result */
    if (rank == 0) {
        x = (float*)malloc(N*sizeof(float));
        for (int i = 0; i < N; i++) {
            x[i] = x_val;
        }
        MPIX_Send_enqueue(x, N, MPI_FLOAT, 1, 0, stream_comm);

        free(x);
    } else if (rank == 1) {
        y = (float*)malloc(N*sizeof(float));
        cudaMalloc(&d_x, N*sizeof(float));
        cudaMalloc(&d_y, N*sizeof(float));

        for (int i = 0; i < N; i++) {
            y[i] = y_val;
        }
        cudaMemcpyAsync(d_y, y, N*sizeof(float), cudaMemcpyHostToDevice, stream);
        MPIX_Recv_enqueue(d_x, N, MPI_FLOAT, 0, 0, stream_comm, MPI_STATUS_IGNORE);
        saxpy<<<(N+255)/256, 256, 0, stream>>>(N, a, d_x, d_y);

        cudaMemcpyAsync(y, d_y, N*sizeof(float), cudaMemcpyDeviceToHost, stream);

        cudaFree(d_x);
        cudaFree(d_y);
        free(y);
    }
    
    MPI_Comm_free(&stream_comm);
    MPIX_Stream_free(&mpi_stream);

    cudaStreamDestroy(stream);
    MPI_Finalize();

    return 0;
}
\end{lstlisting}

\section{Thread Communicators}
\subsection{Background}
MPI is a distributed process model and it requires application designers to 
decompose the problem and parallelize the code from beginning to end. This requires a significant amount of effort.
On the other hand, a typical shared memory-based framework such as OpenMP allows applications to start as a serial program and then add
parallel regions to the computation-intensive part via convenient pragma
directives.
OpenMP is the second most used runtime among HPC applications, next to MPI (\cite{ecp-usage-20}).
For applications outside the HPC community, multithreading is by far the dominant parallelization technique.

However, optimizing a large multithreading application to improve its parallel performance can be challenging.
For example, the ``naive'' loop-level OpenMP pattern will frequently enter and exit parallel regions and incur many implicit
thread synchronizations, easily resulting in poor parallel performance (\cite{krawezik2003}).
To enhance performance,  significant effort is required to enlarge parallel sections.
The extreme version of larger parallel sections is the single program, multiple data pattern,
at which stage the program becomes very close to an MPI-equivalent code.
However, OpenMP does not have APIs as rich as MPI to perform explicit communications, synchronizations, or collective operations.
This forces applications to re-invent code patterns that perform similar MPI tasks between threads.

As the modern node architectures become increasingly
heterogeneous, a growing number of applications adopted a hybrid approach to
parallelization: using OpenMP for on-node parallelization and using MPI
for internode parallelization (\cite{hetero-18}).
This hybrid programming pattern is commonly referred to as MPI+OpenMP, or generally as \MPIpThreads.

In \MPIpThreads, MPI and threads work independently. To acknowledge this
usage, MPI specifies four thread-compatibility levels:  \THREAD{SINGLE},
\THREAD{FUNNELED}, \THREAD{SERIALIZED}, and \THREAD{MULTIPLE}. The levels
merely specify the thread safety of MPI functions; they do not pass the
thread execution context to MPI. The most flexible level, \THREAD{MULTIPLE}, requires
nearly all MPI functions to be thread-safe. In addition, MPI messages are required
to maintain order based on a serial semantic model.
Without explicit thread execution context,  multithreaded MPI communication
generally has low performance, as discussed in the last section.
While the MPIX stream extension addresses the MPI performance issue, it does not address
the complexity that is inherent in a hybrid approach such as \MPIpThreads.

Both OpenMP and MPI share similarities in how both create a parallel environment and provide facilities
for writing parallel codes. While OpenMP focuses on
shared-memory parallelization, and MPI focuses on distributed-memory 
parallelization, both are used for writing similar parallel programming tasks, but with non-interoperable code.
For example, an \MPI{Barrier} synchronizes between processes for the calling thread, whereas an OpenMP barrier
synchronizes only between threads. 
To realize a global barrier among all threads from all processes will require sandwich
calls to both barriers.
On the other hand, OpenMP and MPI provide different sets of APIs that can be complementary.
For example, the dominant feature in OpenMP is dynamic parallelization, which allows a program to enter and exit parallel regions
or even nested parallel regions at will. MPI does offer a dynamic process API, but it is 
difficult to use and is difficult to optimize.
MPI provides rich APIs for explicit communications, synchronizations, and collective operations. Similar functionalities
are often developed \textit{ad hoc} in OpenMP.

\begin{figure}
\tikzstyle{box}=[draw, rectangle, minimum height=0.25cm, fill=yellow]

\resizebox{0.4\textwidth}{!}{
\begin{tikzpicture}
\draw (0, 0) -- +(0.5, 0);
\node[fill=white] at (-0.5,0) {\ttfamily main};

\node[box, minimum width=1.2cm, fill=yellow] at (1.5, -0.5) {};
\draw[thin] (0.5,0)--(0.9, -0.5);
\draw[thin] (2.1,-0.5)--(2.5, 0);
\node[box, minimum width=1.2cm, fill=yellow] at (1.5, 0) {};
\draw[thin] (0.5,0)--(0.9, 0);
\draw[thin] (2.1,0)--(2.5, 0);
\node[box, minimum width=1.2cm, fill=yellow] at (1.5, 0.5) {};
\draw[thin] (0.5,0)--(0.9, 0.5);
\draw[thin] (2.1,0.5)--(2.5, 0);
\draw (2.5, 0) -- +(0.5, 0);

\node[box, minimum width=1.2cm, fill=yellow] at (4, -0.75) {};
\draw[thin] (3,0)--(3.4, -0.75);
\draw[thin] (4.6,-0.75)--(5, 0);
\node[box, minimum width=1.2cm, fill=yellow] at (4, -0.25) {};
\draw[thin] (3,0)--(3.4, -0.25);
\draw[thin] (4.6,-0.25)--(5, 0);
\node[box, minimum width=1.2cm, fill=yellow] at (4, 0.25) {};
\draw[thin] (3,0)--(3.4, 0.25);
\draw[thin] (4.6,0.25)--(5, 0);
\node[box, minimum width=1.2cm, fill=yellow] at (4, 0.75) {};
\draw[thin] (3,0)--(3.4, 0.75);
\draw[thin] (4.6,0.75)--(5, 0);
\draw (5, 0) -- +(0.5, 0);

\node[box, minimum width=1.2cm, fill=yellow] at (6.5, -0.5) {};
\draw[thin] (5.5,0)--(5.9, -0.5);
\draw[thin] (7.1,-0.5)--(7.5, 0);
\node[box, minimum width=1.2cm, fill=yellow] at (6.5, 0) {};
\draw[thin] (5.5,0)--(5.9, 0);
\draw[thin] (7.1,0)--(7.5, 0);
\node[box, minimum width=1.2cm, fill=yellow] at (6.5, 0.5) {};
\draw[thin] (5.5,0)--(5.9, 0.5);
\draw[thin] (7.1,0.5)--(7.5, 0);
\draw (7.5, 0) -- +(0.5, 0);

\draw[dashed, very thick, rounded corners=0.2cm] (3.5, 1) rectangle (4.5, -1);
\draw[->] (6,1.5) node[fill=white] {\rmfamily thread communicator} -- (4.5, 1.1);

\end{tikzpicture}
}
\caption{Diagrams illustrating the parallel programming patterns using thread communicator.}
\label{fig:threadcomm}
\end{figure}
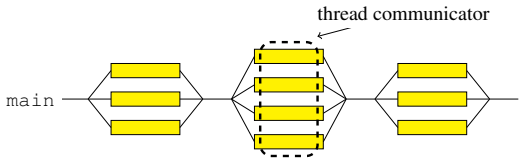

So instead of the hybrid approach in \MPIpThreads, can we extend MPI and allow it to be
used inside a parallel region directly between threads as well as between processes (illustrated in Fig~\ref{fig:threadcomm})?
In an $N$-process MPI program and an $M$-thread OpenMP parallel region, our proposed extension will create
an MPI communicator of size $N \times M$.
This approach would create a complementary way of using MPI and OpenMP. MPI can utilize OpenMP's
flexibility in creating dynamic parallel regions and expand its parallelization. OpenMP, on the
other hand, can use MPI's explicit messaging, and collective APIs to achieve
cleaner code and better performance.
This new programming pattern is referred to as \MPIxThreads \hskip .05in (\cite{mpixthreads}), and it is implemented in MPICH as an extension called thread communicator.

\subsection{Extension}
\begin{lstlisting}[style=proto]
int MPIX_Threadcomm_init(MPI_Comm comm, int num_threads,
                         MPI_Comm *newthreadcomm)

int MPIX_Threadcomm_free(MPI_Comm *threadcomm)
\end{lstlisting}
These two functions are called outside thread-parallel regions, such as code marked by OpenMP pragama \texttt{omp single}. \texttt{MPIX\_Threadcomm\_init} creates a thread communicator. Its semantics are similar to \texttt{MPI\_Comm\_dup} with an additional attribute setting the number of threads the communicator will be used among. Different processes may specify different numbers of threads.
Outside a thread-parallel region, the thread communicator is inactive and it cannot be used for MPI communications. It only can be activated inside a parallel region, and it can be activated and deactivated multiple times. Once the thread communicator is no longer needed, it should be freed with \texttt{MPIX\_Threadcomm\_free}.
\begin{lstlisting}[style=proto]
int MPIX_Threadcomm_start(MPI_Comm threadcomm)

int MPIX_Threadcomm_finish(MPI_Comm threadcomm)
\end{lstlisting}
These two functions are used inside a thread-parallel region to activate and deactivate a thread communicator. The exact number of threads as specified in \texttt{MPIX\_Threadcomm\_init} need to call these functions collectively. After the thread communicator is activated, each thread is assigned a unique rank, and can call MPI operations as if it is an MPI process.

\begin{lstlisting}[style=proto]
int MPIX_Comm_test_threadcomm(MPI_Comm comm, int *flag)
\end{lstlisting}
This function can be used by portable codes to reliably tell whether a communicator is a thread communicator or a conventional communicator.

\subsection{Example}
The following example shows the usage of the thread communicator extension.
Once the thread communicator is activated inside a parallel region, regular MPI functions can be used for interthread synchronization. They are omitted here for brevity.
\begin{lstlisting}[style=example]
#include <mpi.h>
#include <stdio.h>
#include <assert.h>

#define NT 4

int main(void) {
    MPI_Comm threadcomm;

    MPI_Init(NULL, NULL);
    MPI_Threadcomm_init(MPI_COMM_WORLD, NT,
                        &threadcomm);

    #pragma omp parallel num_threads(NT)
    {
        assert(omp_get_num_threads() == NT);
        int rank, size;
        MPI_Threadcomm_start(threadcomm);
        MPI_Comm_size(threadcomm, &size);
        MPI_Comm_rank(threadcomm, &rank);
        printf("  Rank %d / %d\\n", rank, size);

        /* MPI operations over threadcomm */

        MPI_Threadcomm_finish(threadcomm);
    }

    MPI_Threadcomm_free(&threadcomm);
    MPI_Finalize();
    return 0;
}
\end{lstlisting}

Running the example, we show that each thread within the OpenMP parallel region behaves as an MPI process.
\begin{lstlisting}[style=example, language=sh, frame=tb]
$ mpicc -fopenmp -o t t.c
$ mpirun -n 2 ./t
    Rank 4 / 8
    Rank 7 / 8
    Rank 5 / 8
    Rank 6 / 8
    Rank 0 / 8
    Rank 1 / 8
    Rank 2 / 8
    Rank 3 / 8
\end{lstlisting}

\subsection{Evaluation}
\begin{figure}[htp]
    \centering
    \includegraphics[clip,width=\columnwidth]{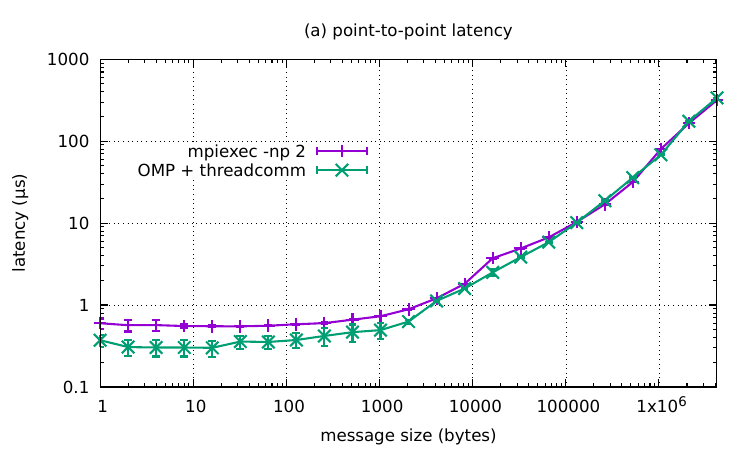}
    \includegraphics[clip,width=\columnwidth]{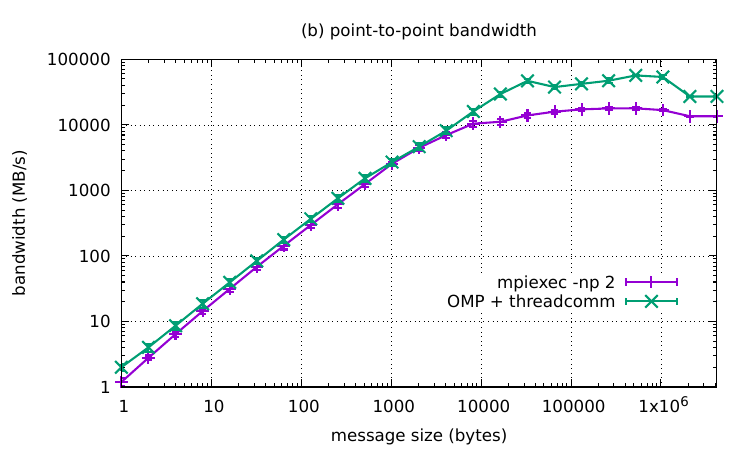}
    \caption{Point-to-point message latency and bandwidth comparison between MPI-everywhere and OpenMP+threadcomm on an Intel Xeon Gold 5317. Processes or threads are bound to cores on the same socket.}
    \label{fig:omp_latency}
\end{figure}

The conventional way of using MPI for on-node programming is to launch multiple MPI processes on a single node. The latter is referred to as MPI-everywhere.
The thread communicator extension allows such MPI programs to be ported to OpenMP without modifying the bulk of the MPI code.
In Figure~\ref{fig:omp_latency} we compare the performance of equivalent MPI point-to-point messaging codes using OpenMP plus thread
communicator versus MPI-everywhere.

The point-to-point communication latency results are shown in Figure~\ref{fig:omp_latency}(a) and bandwidth results shown in Figure~\ref{fig:omp_latency}(b).
Both show similar performance as expected since both are implemented as shared-memory-based communication. 
The interthread messaging via threadcomm shows slightly better performance, especially for small message latency and large message bandwidth.
The reduced latency for small messages stems from a shortcut that bypasses the allocation and deallocation of sender request objects.
Request objects are necessary when a message exceeds the threshold of pre-allocated message cells, necessitating asynchronous tracking. 
By omitting the request object for small messages, the instruction count decreases, thereby enhancing latency. While this optimization could theoretically extend to interprocess shared-memory messaging, its implementation is more challenging due to constraints within the current MPICH framework.
For larger messages, interthread messaging employs the single-copy algorithm, while interprocess messaging utilizes the two-copy rendezvous algorithm.
We observe a decline in bandwidth beyond the 1 MB message size, which we attribute to heavy last-level cache and TLB misses.

The better performance of threadcomm interthread messaging demonstrates that deploying one process per compute node and utilizing OpenMP to initiate parallel regions serves as a highly effective alternative to the MPI-everywhere model.

\section{General Progress}
\subsection{Background}
Overlapping communication and computation is an important technique to obtain high performance in high-performance computing.
MPI offers nonblocking routines that split an otherwise blocking communication call into two parts, the starting call and the wait/test completion call.
However, the actual computation/communication overlap relies on the communication progress happening between the starting call and the completion call.
This is not always true. It depends on the underlying communication system and the implementation algorithms.
For example, if a large message is split into chunks transmitted via pipelining, frequent calls into MPI progress (via MPI\_Wait/MPI\_Test) may be necessary to acknowledge a message chunk and start the next chunk. If the application goes into computation after the MPI\_Isend/MPI\_Irecv call and only calls MPI\_Wait after the computation completes, the actual transmission of the message may be delayed without the computation/communication overlap. This is illustrated in Figure~\ref{fig:progress}(a).
To achieve overlapping, one must invoke MPI progress during computation or spawn a progress thread that invokes progress in the background, as illustrated in Figure~\ref{fig:progress}(b).

MPI also offers one-sided communication (also known as remote memory access or RMA) that allows one process to specify all communication parameters for both the origin and target sides. While the target side is not semantically involved in the RMA operations, for many systems and implementations it is required to participate in the progress or the data transmission may be delayed. Thus, making background progress can be a key in solving the RMA performance issue.

With MPICH, a background progress thread can be added simply by setting an environment variable \texttt{MPIR\_CVAR\_ASYNC\_PROGRESS}. While it is convenient, launching a general-purpose background progress thread as such has two major drawbacks.
First, the progress thread has to invoke or poll progress frequently or it may increase the message latency. The busy polling in the progress thread will take a CPU core away that otherwise can be used for computation. This is notably undesirable when applications launch multiple processes for each node, which is common on today's many-core systems.
Second, the progress thread forces the application to run in \THREAD{MULTIPLE}, and the thread will constantly compete for critical sections with every other thread that is making MPI calls. As we have seen earlier, this will significantly degrade the communication performance.
While the drawbacks of asynchronous progress are difficult for implementations to address, they are simple for applications to control. In principle, an application can spin up the background progress when it requires background progress and when messaging latency is critical. It can spin down the background progress, either by inserting sleeps between the progress poll or completely pausing the progress, when messaging latency is not critical or background progress is not needed.
In addition, by using MPIX stream, applications can separate the communication traffic into different channels and make background progress on specific channels, thus avoiding performance degradation as we have seen in \THREAD{MULTIPLE}.

MPICH provides extensions to allow applications to construct custom asynchronous progress threads.

\begin{figure}
\tikzstyle{blkcpu}=[draw, rectangle, minimum height=0.5cm, fill=red, anchor=west]
\tikzstyle{blknic}=[draw, rectangle, minimum height=0.5cm, fill=blue, text=white, anchor=west]
\tikzstyle{blkcomp}=[draw, rectangle, minimum height=0.5cm, fill=green, anchor=west]
\tikzstyle{blkbg}=[draw, rectangle, minimum height=0.5cm, fill=black!15, anchor=west]
\tikzstyle{txt}=[anchor=west]

\resizebox{0.4\textwidth}{!}{
\begin{tikzpicture}
\node[txt] at (0.4, 0.75) {\small Isend/Irecv};

\node[blkcpu, minimum width=0.3cm] at (0,0) {};
\node[blkcomp, minimum width=3.0cm] at (0.3, 0) {\small Computation};
\node[blkcpu, minimum width=2.6cm] at (3.3, 0) {\small Wait};
\node[blknic, minimum width=2.5cm] at (3.3, -1) {\small Transmission};

\draw[->] (0.4, 0.6) -- (0.2, 0.35);

\draw (0, 1) -- (0, -1.5);
\draw[->, very thick] (0, -1.5) -- (6, -1.5);
\node at(5.8, -1.8) {Time};

\node at(3, -2.5) {\scriptsize (a) No progress during computation.};


\node[blkcpu, minimum width=0.3cm] at (0,-4) {};
\node[blkcomp, minimum width=3.0cm] at (0.3, -4) {\small Computation};
\node[blkcpu, minimum width=0.15cm] at (3.3, -4) {};

\node[blknic, minimum width=2.5cm] at (0.3, -4-1) {\small Transmission};
\node[blkbg, minimum width=5.5cm] at (0, -4-2) {\small background progress};

\node[txt] at (0.4, -4+0.75) {\small Isend/Irecv};
\node[txt] at (3.5, -4+0.75) {\small Wait};
\draw[->] (0.4, -4+0.6) -- (0.2, -4+0.35);
\draw[->] (3.7, -4+0.6) -- (3.5, -4+0.35);

\draw (0, -4+1) -- (0, -4-2.5);
\draw[->, very thick] (0, -4-2.5) -- (6, -4-2.5);
\node at(5.8, -4-2.8) {Time};

\node at(3, -4-3.5) {\scriptsize (b) With background progress.};
\end{tikzpicture}
}
\caption{Diagrams illustrating asynchronous progress.}
\label{fig:progress}
\end{figure}

\subsection{Extension}
The standard way of invoking MPI progress is via \MPI{Test} (as well as \texttt{MPI\_Test\{any,some,all\}}) or the blocking \MPI{Wait}. These MPI completion calls require specific MPI request handles. But for the purpose of making general asynchronous progress, it is unnecessary and sometimes it is impossible to tie the progress to a specific request handle. Thus, we need a general progress invoking function. 
\begin{lstlisting}[style=proto]
int MPIX_Stream_progress(MPIX_Stream stream)
\end{lstlisting}
This function invokes progress on a given MPIX stream. \texttt{MPIX\_STREAM\_NULL} can be used to invoke general progress on all implicit streams.

We recommend that applications spawn their own threads to poll \texttt{MPIX\_Stream\_progress} so that they can control the polling frequency and be able to spin up or down the progress. For convenience, MPICH also provides the following routines to start and stop a default progress thread on a given MPIX stream.
\begin{lstlisting}[style=proto]
int MPIX_Start_progress_thread(MPIX_Stream stream);

int MPIX_Stop_progress_thread(MPIX_Stream stream);
\end{lstlisting}

\subsection{Example}
The following example shows an implementation of a background progress thread using \texttt{MPIX\_Stream\_progress}. 
The process with \texttt{origin\_rank} issues a series of RMA operations with a passive synchronization. However, many MPI implementations require progress at the target process for passive synchronization or the RMA operations will get delayed. In the example, if we omit the background progress thread at the target process, it will take the duration of the target process being busy before the RMA operations can be completed. By enabling the progress thread at the target process, the RMA operations are completed immediately. The background thread makes progress by calling \texttt{MPIX\_Stream\_progress}. A volatile flag variable is used to control the background thread to make busy progress only when it is needed.
\begin{lstlisting}[style=example]
/* progress.c - run with at least 2 processes */
#include <mpi.h>
#include <stdio.h>
#include <unistd.h>
#include <pthread.h>
#include <assert.h>

#define MAX_DATA_SIZE 1024

int buf[MAX_DATA_SIZE];
int win_buf[MAX_DATA_SIZE];

enum {
    PROGRESS_IDLE,
    PROGRESS_BUSY,
    PROGRESS_EXIT,
};
volatile int need_progress = PROGRESS_IDLE;

void *progress_thread(void *ptr)
{
    while(1) {
        if (need_progress == PROGRESS_IDLE) {
            sleep(1);
        } else if (need_progress == PROGRESS_BUSY) {
            MPIX_Stream_progress(MPIX_STREAM_NULL);
        } else if (need_progress == PROGRESS_EXIT) {
            break;
        }
    }
}

int main(void)
{
    int thread_level;
    MPI_Init_thread(NULL, NULL, MPI_THREAD_MULTIPLE, &thread_level);
    assert(thread_level == MPI_THREAD_MULTIPLE);

    int rank, size;
    MPI_Comm_rank(MPI_COMM_WORLD, &rank);

    for (int i = 0; i < MAX_DATA_SIZE; i++) {
        win_buf[i] = i;
    }

    int origin_rank = 0;
    int target_rank = 1;
    pthread_t thread;
    if (rank == target_rank) {
        pthread_create(&thread, NULL, progress_thread, NULL);
    }

    MPI_Win win;
    MPI_Win_create(win_buf, MAX_DATA_SIZE, 4, MPI_INFO_NULL, MPI_COMM_WORLD, &win);
    if (rank == origin_rank) {
        double time_start = MPI_Wtime();
        MPI_Win_lock(MPI_LOCK_SHARED, target_rank, 0, win);
        for (int i = 0; i < MAX_DATA_SIZE; i++) {
            MPI_Get(buf + i, 1, MPI_INT, target_rank, i, 1, MPI_INT, win);
        }
        MPI_Win_unlock(target_rank, win);
        printf("Completed all gets in %.3f seconds\n", MPI_Wtime() - time_start);
    } else if (rank == target_rank) {
        /* If we do not start background progress, the passive RMA in rank 0 won't complete while the target rank is busy */
        need_progress = PROGRESS_BUSY;
        /* Sleep to simulate busy computations */
        sleep(10);
    }

    MPI_Barrier(MPI_COMM_WORLD);
    if (rank == target_rank) {
        need_progress = PROGRESS_EXIT;
    }

    MPI_Win_free(&win);

    void *ret;
    pthread_join(thread, &ret);

    MPI_Finalize();

    return 0;
}


\end{lstlisting}

\section{Summary}
In this paper, we document the nonstandard MPI extensions that are included in MPICH as of version 4.2.0.
The generalized requests extension allows applications to integrate asynchronous tasks with MPI requests and MPI progress without extra threads.
The datatype iovec extension allows applications to use MPI derived datatypes to describe general memory layout and apply them to usages in addition to MPI communications.
The MPIX stream extension allows applications to explicitly map execution contexts to MPI's internal streams such as the virtual communication contexts to achieve optimum parallel performances.
The stream enqueue extension allows applications to place MPI communications onto an offloading context such as  GPU accelerators.
The thread communicator extension allows MPI APIs to be used between threads in thread parallel regions, thus providing a more consistent programming environment for MPI+Threads applications.
The MPI stream progress extension allows applications to invoke general progress without tying to a particular MPI communication, thus allowing it to spawn custom progress threads that minimally interfere with the main thread.

All these extensions extend various parts of MPI, such as MPI requests, MPI datatypes, MPI execution contexts, and MPI progress, to be more interoperable in an increasingly more hybrid computing environment.

\begin{acks}
    We gratefully acknowledge the computing resources provided and operated by the
Joint Laboratory for System Evaluation (JLSE) at Argonne National Laboratory.
This research was supported by the Exascale Computing Project (17-SC-20-SC),
a collaborative effort of the U.S.\ Department of Energy Office of Science and
the National Nuclear Security Administration, and the U.S. Department of
Energy, Office of Science, under Contract DE-AC02-06CH11357.
\end{acks}

\bibliographystyle{SageH}
\bibliography{references}

\end{document}